\documentclass[prd,reprint,onecolumn,linenumbers,	
amssymb, amsmath, aps, showpacs, nofootinbib, superscriptaddress]{revtex4}

\usepackage{graphicx}
\usepackage{dcolumn}
\usepackage{bm}

\newcommand{\beq}{\begin{equation}}
\newcommand{\eeq}{\end{equation}}
\newcommand{\beqa}{\begin{eqnarray}}
\newcommand{\eeqa}{\end{eqnarray}}

\begin{document}

\title{Fast radio bursts as a cosmic probe?}
\affiliation{Key Laboratory of Dark Matter and Space Astronomy, Purple Mountain Observatory, Chinese Academy of Sciences, Nanjing 210008, China}
\affiliation{Graduate University of Chinese Academy of Sciences, Yuquan Road 19, Beijing, 100049, China.}
\affiliation{Laboratoire AIM-Paris-Saclay, CEA/DSM/Irfu, CNRS, Universite Paris Diderot, Saclay, pt courrier 131, 91191 Gif-sur-Yvette, France}
\affiliation{School of Astronomy and Space Science, Nanjing University, Nanjing 210093, China}
\affiliation{Key Laboratory of Modern Astronomy and Astrophysics (Nanjing University)£¬Ministry of Education, Nanjing 210093, China}
\author{Bei Zhou}
\affiliation{Key Laboratory of Dark Matter and Space Astronomy, Purple Mountain Observatory, Chinese Academy of Sciences, Nanjing 210008, China}
\affiliation{Graduate University of Chinese Academy of Sciences, Yuquan Road 19, Beijing, 100049, China.}
\author{Xiang Li}
\affiliation{Key Laboratory of Dark Matter and Space Astronomy, Purple Mountain Observatory, Chinese Academy of Sciences, Nanjing 210008, China}
\affiliation{Graduate University of Chinese Academy of Sciences, Yuquan Road 19, Beijing, 100049, China.}
\author{Tao Wang}
\affiliation{Laboratoire AIM-Paris-Saclay, CEA/DSM/Irfu, CNRS, Universite Paris Diderot, Saclay, pt courrier 131, 91191 Gif-sur-Yvette, France}
\affiliation{School of Astronomy and Space Science, Nanjing University, Nanjing 210093, China}
\affiliation{Key Laboratory of Modern Astronomy and Astrophysics (Nanjing University)£¬Ministry of Education, Nanjing 210093, China}
\author{Yi-Zhong Fan$^\ast$}
\affiliation{Key Laboratory of Dark Matter and Space Astronomy, Purple Mountain Observatory, Chinese Academy of Sciences, Nanjing 210008, China}
\author{Da-Ming Wei}
\affiliation{Key Laboratory of Dark Matter and Space Astronomy, Purple Mountain Observatory, Chinese Academy of Sciences, Nanjing 210008, China}

\date{\today}

\begin{abstract}
We discuss the possibility of using fast radio bursts (FRBs), if cosmological, as a viable cosmic probe.
We find out that the contribution of the host galaxies to the detected dispersion measures can be inapparent for the FRBs not from galaxy centers or star forming regions.
The inhomogeneity of the intergalactic medium (IGM), however, causes significant deviation of the dispersion measure from that predicted in the simplified homogeneous IGM model for individual event.
Fortunately, with sufficient FRBs along different sightlines but within a very narrow redshift interval (e.g., $\Delta z \sim 0.05$ or $\Delta z \sim 0.05(1+z)$), the mean from averaging observed dispersion measures does not suffer such a problem and hence may be used as a cosmic probe.
We show that in the optimistic case (e.g., tens FRBs in each $\Delta z$ have been measured; the most distant FRBs were at redshift $\geq 3$; the host galaxies and the FRB sources contribute little to the detected dispersion measures) and with all the uncertainties (i.e. the inhomogeneity of the IGM, the contribution and uncertainty of host galaxies as well as the evolution and error of $f_{\rm IGM}$) considered,
FRBs could help constrain the equation of state of dark energy.
\end{abstract}

\pacs{98.70.Dk, 98.70.-f, 98.80.-k}

\maketitle

Very recently, Thornton et al. \cite{Thornton2013} have reported the detection of four millisecond-duration radio bursts (hereafter FRBs) all more than $40^{\circ}$ from the Galactic plane, confirming Lorimer et al. \cite{Lorimer2007}'s discovery. Current data
favor celestial rather than terrestrial origin since the arrival times of individual pulses follow the $\nu^{-2}$ law that characterizes the propagation of radio waves through a cold plasma, where $\nu$ is the observer's frequency. The host galaxy and
intergalactic medium models suggest that the FRBs have cosmological redshifts ($z$) of 0.5 to 1 and distances
of up to $\sim 3$ Gpc. On the other hand, so far no reliable $\gamma$-ray, X-ray or optical counterparts have been identified and only tentative association of single dispersed millisecond radio pulses with two Gamma-ray Bursts has been reported \cite{Bannister2012}. Now FRBs have attracted wide attention and some groups have started to do the ``followup" observations. If counterparts can be reliably identified and the cosmological origin has been directly confirmed, one interesting question people would ask is whether FRBs can help in measuring the geometry of the universe. This is because the dispersion measures (DMs) of the radio bursts contain valuable information on the cosmological distance they have traveled, which in turn provides us the chance to constrain the parameters governing the universe particularly if some FRBs were at redshifts $z\sim$ quite a few. In the sample of \cite{Thornton2013}, the brightest one is FRB 110220 with a flux $\sim 2.5$ Jy (at 1.5 GHz) and the inferred redshift is $z\sim 0.81$. The other 3 FRBs are dimmer by a factor of $\sim 3-4$. Due to the narrow band of the detector, the energy spectrum of FRBs is still unknown. If the spectrum in the range of 1-10 GHz is as hard as (or harder than) that of FRB 110220 in the range of 1.219-1.494 GHz, events similar to FRB 110220 but at $z\sim 3$ are observable. Motivated by these facts, we discuss the prospect of using FRBs as a new cosmic probe.
In particular, we investigate their potential in better probing the equation state of the dark energy since in some models the equation-of-state parameter $w$ differs from $-1$ and evolves with $z$ \cite{Copeland2006}.
We are aware that the possible roles of FRBs in probing the (missing) baryons in the intergalactic medium (IGM) have been discussed in \cite{McQuinn2014,Deng2014}.

In general, the rate of expansion of the universe at a redshift $z$, $H(z)$, is described by the
Friedmann equation
\begin{equation}
{H^{2}(z)\over H_0^{2}}=\Omega_{\rm M}(1+z)^{3}+\Omega_{\rm DE}\exp{\int 3(1+w(z))d\ln(1+z)}+\Omega_{\rm k}(1+z)^{2},
\end{equation}
where $H_0$ is the rate of expansion today, $\Omega_{\rm M}$ and $\Omega_{\rm DE}$ are the
matter (including baryons and dark matter) and dark-energy density with respect to the critical density
$\rho_{\rm c}=3c^{2}H_0^{2}/8\pi G$
today,
and $\Omega_{\rm k}=1-\Omega_{\rm M}-\Omega_{\rm DE}$ is the spatial curvature density of the
universe \cite[e.g.,][]{Suzuki2012}, $G$ is Newton's constant. Distances, such as the distance traveled by the photons, depend on
the integral of $1/H(z)$ over redshift (i.e., $dL_{\rm ph}=cdz/[(1+z)H(z)]$). As a preliminary approach, we simply take $\Omega_{\rm k}=0$ and $w={\rm const.}$
The mean dispersion measure caused by the inhomogeneous intergalactic medium is given by
\begin{equation}
\langle{\rm DM}_{\rm IGM}\rangle(z) =\Omega_{\rm b} \frac{3 H_0 c}{8\pi G m_{\rm p}}\int_{0}^{z} {(1+z')f_{\rm IGM}(z')(Y_{\rm H} X_{\rm e,H}(z')+\frac{1}{2} Y_{\rm p} X_{\rm e,He}(z')) \over \{\Omega_{\rm M}(1+z')^{3}+\Omega_{\rm DE}(1+z')^{3[1+w(z')]}\}^{1/2}}dz',
    \label{eq:DM_IGM}
\end{equation}
where $X_{\rm e,H}$ and $X_{\rm e,He}$ are the ionization fractions of the intergalactic hydrogen and helium, respectively, and $Y_{\rm H}=\frac{3}{4}$, $Y_{\rm p}=\frac{1}{4}$ is mass fraction of H, He.
Different from \cite{Ioka2003,Deng2014}, we call the above expression the ``mean dispersion measure" since the IGM is found to be inhomogeneous and significant fluctuation of individual measurements along different sightlines is expected \cite{McQuinn2014}. As shown later, such a fluctuation has to be carefully addressed in our modeling. Current observations suggest that for $z\lesssim 6$ ($\lesssim 3$) the intergalactic hydrogens (helium) are fully ionized \cite{Meiksin2009,Becker2011}. For $z\lesssim 3$, it is reasonable to take $X_{\rm e,H}=1$ and $X_{\rm e,He}=1$.
At $z\gtrsim 1.5$, some 90\% of the baryons produced in the Big Bang are contained within the IGM (i.e., $f_{\rm IGM}\approx 90\%$), with only 10\% in galaxies, galaxy clusters or possibly locked up in an early generation of compact stars \cite{Meiksin2009}.
While at redshifts $z\leq 0.4$, the locked baryons are found to account for $\approx 18\pm4\%$ of the total \cite{Shull2012} and we have $f_{\rm IGM}\approx 82\%$. The slowly evolving $f_{\rm IGM}$ at low redshifts is indeed a challenge for our purpose. However, as many FRBs with known redshifts and DMs have been collected, with proper treatments the influence of the evolving $f_{\rm IGM}$ in constraining cosmological parameters might be effectively suppressed. For example, if we adopt $\Delta\langle{\rm DM}\rangle=\langle{\rm DM}\rangle(z)-\langle{\rm DM}\rangle(z_{\rm b})$ for $z_{\rm b}\sim 1$ rather than $\langle{\rm DM}\rangle(z)$ to constrain the cosmological parameters, the effect of evolving $f_{\rm IGM}$ at low redshifts would be reasonably removed.
Therefore in the rest of this work we take $f_{\rm IGM}=0.9$ at ${z\geq1.5}$, while at $z<1.5$, we take $f_{\rm IGM}$ to be increased linearly from $0.82$ at $z=0$ to 0.9 at $z=1.5$ and a random deviation of $0.04$ is assumed.

The type Ia SNe are excellent cosmology probe \cite{Riess1998} and the most distant Type Ia SN yet observed was at the redshift $z=1.914$ \cite{Jones2013}. To be a viable/complementary cosmic probe, $\langle {\rm DM} \rangle(z)$ should be obtained within an accuracy of $\sim 10\%$, i.e., comparable with the accuracy of the luminosity distance measurement with SN Ia \cite{Suzuki2012}. The observed dispersion measure of each FRB consists of four components, including that of the Milky Way (${\rm DM}_{\rm MW}$), the intergalactic medium (${\rm DM}_{\rm IGM}$), the host galaxies (${\rm DM}_{\rm host}$) and the FRB sources (${\rm DM}_{\rm sour}$). In the lack of a reasonable understanding of the physical origin of FRBs, it is not possible to estimate ${\rm DM}_{\rm sour}$ reliably (see \cite{Deng2014} for the discussion within a specific model). Thornton et al. \cite{Thornton2013} took the sum of ${\rm DM}_{\rm sour}$ and ${\rm DM}_{\rm host}$ to be $100~{\rm pc}~{\rm cm^{-3}}$. Similarly we assume that ${\rm DM}_{\rm sour}$ is ignorable. The intervening galaxies, if there are, will make the case more complicated. With sufficient FRBs with identified counterparts, the events with intervening galaxy (galaxies) can be excluded without losing significant information. The dispersion measures of Milky Way galaxy along different sightlines have been extensively discussed \cite{Schnitzeler2012}, which can be reasonably removed from the data. That is why in the following discussion we focus on ${\rm DM}_{\rm IGM}$ and ${\rm DM}_{\rm host}$.

\begin{figure}
\includegraphics[width=90mm,angle=0]{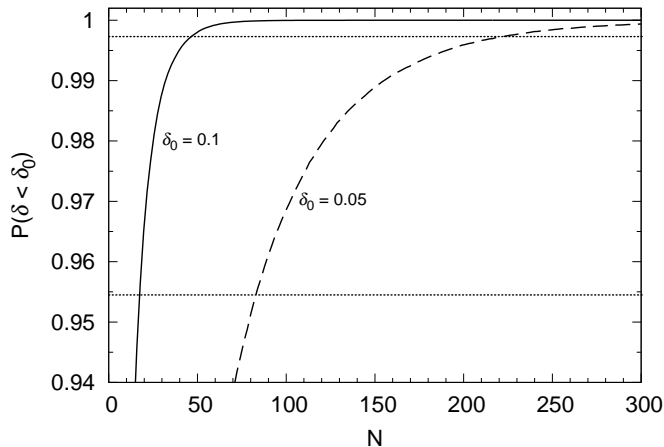}
\caption{The possibility to get a $\delta<\delta_0$ in averaging the inferred dispersion measures as a function of $N$, the number of events with redshifts ranging from $1$ to $1.05$. The probability distribution function of dispersion measures used in our Monte Carlo simulation is adopted from \cite{McQuinn2014} (i.e., the solid line plotted in the bottom panel of Fig.2 therein).
}
\label{fig:deviation}
\end{figure}

As pointed out in \cite{McQuinn2014}, for individual FRBs, the detected dispersion measures ${\rm DM}_{\rm IGM}(z)$ may remarkably differ from $\langle {\rm DM}_{\rm IGM}\rangle(z)$ given in eq.(\ref{eq:DM_IGM}). For example, at $z\sim1$, the sightline-to-sightline variance in ${\rm DM}_{\rm IGM}(z)$ (i.e., $\sigma_{\rm DM_{\rm IGM}}(z)$) is expected to be $\sim 180-400~{\rm pc~cm^{-3}}$, depending on the halo's gas profiles of the ionized baryons (see Fig.1 of \cite{McQuinn2014}). Correspondingly, the ratio $\sigma_{\rm DM_{\rm IGM}}/\langle {\rm DM}_{\rm IGM}\rangle$ is within the range of  $\sim 16\%-35\%$. For $z>1.1$, the increase of $\sigma_{\rm DM_{\rm IGM}}(z)$ is only moderate since the high redshift universe is becoming more homogenous. Nevertheless, at $z\sim 1.5$, one still has $\sigma_{\rm DM_{\rm IGM}}/\langle {\rm DM}_{\rm IGM} \rangle \sim 12\%-27\%$. Such a large $\sigma_{\rm DM_{\rm IGM}}/\langle {\rm DM}_{\rm IGM} \rangle$ is a serious challenge for using FRBs as a viable cosmic probe. Fortunately, if there are sufficient FRBs from different sightlines but at very similar redshifts (for example, $\Delta z \sim 0.05$), their mean $\overline{\rm DM}_{\rm IGM}(z)$ would be a reasonable approximation of $\langle {\rm DM}_{\rm IGM}\rangle(z)$. We denote the relative deviation as $\delta(z) \equiv |\overline{\rm DM}_{\rm IGM}(z)-\langle {\rm DM_{\rm IGM}}\rangle(z)|/\langle {\rm DM_{\rm IGM}}\rangle(z)$. With a given probability distribution function of dispersion measures, we carry out the Monte Carlo simulation to estimate $\delta$ (The method used in this work is similar to that of \cite{Feng2013}). In Fig.\ref{fig:deviation} we present the possibility of getting a $\delta < \delta_0$ in averaging the observed dispersion measures as a function of $N$, the number of events in the redshift range of $1-1.05$.  At the 95.4\% confidence level, one needs $N \sim 20$ for $\delta_0 =0.1$ and $N\sim 80$ for $\delta_0 =0.05$. One would need $N \sim 45$ for $\delta_0 =0.1$ and $N\sim 225$ for $\delta_0 =0.05$ to reach the 99.7\% confidence level. Therefore $\sim 10^{3}$ events with directly-measured $z$ and DMs are needed for constraining the cosmological parameters reliably with FRBs.
Essentially $z$ can be measured in two ways: either directly observe the ``afterglow" of the FRBs or observe their associated host galaxy. With current facilities, direct spectroscopic measurements of the afterglow are challenging due to i) the position of FRBs are poorly constrained due to the large beam of single-dish telescopes; ii) the afterglow is likely faint, as suggested in some cosmological models \cite{Totani2013}. However, with the upcoming radio transient surveys conducted by radio interferometers, e.g., Square Kilometer Array (SKA, https://www.skatelescope.org, started at 2018), we expect to discover and precisely localize a large number of FRBs. With its wider field of view and higher sensitivity, the SKA will be able to survey the sky at a rate faster than any current telescopes. On the other hand, starting at the beginning of the next decade, the Large Synoptic Survey Telescope (LSST) system will produce a 6-band wide filed survey of roughly  half of the sky using an 8.4-meter telescope. The survey is capable of detecting typical star-forming galaxies ($M_{\rm B}\gtrsim -21$) out to $z >5.5$, and is expected to observe a total number of $10^{10}$ galaxies. The combination of LSST and SKA may allow direct observations of the afterglow at low redshifts and efficient identifications of the host galaxy of FRBs up to $z\gtrsim 5$.
With a high rate $R_{\rm FRB} \sim 10^{4}~{\rm sky^{-1}}~{\rm day^{-1}}$, in the foreseeable future people may be able to have a large sample containing $\sim 10^{3}$ FRBs with DM and redshift measurements then render the approach suggested here realizable.

The other problem is whether ${\rm DM}_{\rm host}$ can be ignored or not. For a host galaxy at a redshift $z$, its contribution to the observed DM is a factor of $1/(1+z)$ of the local one (${\rm DM_{\rm host,loc}}$) as the results of the cosmological time dilation and the frequency shift. So the fraction of DM contributed by the host galaxy is
\begin{equation}
{\cal R}\approx {\rm DM_{\rm host}\over DM_{\rm host}+\langle DM_{\rm IGM}\rangle}\approx  {\rm DM_{\rm host,loc}\over DM_{\rm host,loc}+(1+z)\langle DM_{\rm IGM}\rangle}.
\label{eq:ratio}
\end{equation}
Since $\langle{\rm DM}_{\rm IGM}\rangle$ increases with $z$ quickly (see the thick solid line in Fig.1 of \cite{McQuinn2014}), if ${\rm DM_{\rm host,loc}}$ does not and is smaller than ${\rm a~few}\times 100~{\rm pc~cm^{-3}}$, the using of the ``high redshift" FRBs with $z\sim$ a few as a cosmic probe will be promising. The question is in reality how large ${\rm DM_{\rm host,loc}}$ is. For the Milky Way galaxy, the line of sight passing through the central regions of Milky Way galaxy could lead to DMs in
excess of $700~{\rm  pc~cm^{-3}}$ in the central 100 pc \cite{Deneva2009},
independent of the line-of-sight inclination. With such a huge ${\rm DM_{\rm host,loc}}$ only the FRBs at $z>2$ can be used as viable cosmic probe (i.e., ${\cal R}\leq 10\%$). Fortunately, the DMs along the high-latitude sightlines at other sites, for example the sun, can be much lower (i.e., $\leq 100~{\rm pc~cm^{-3}}$).
The FRBs with ${\rm DM_{\rm host,loc}}<100~{\rm pc~cm^{-3}}$ at $z>1$ are highly needed for our purpose. For the cosmological galaxies, ${\rm DM_{\rm host,loc}}$ is poorly known. DM$_{\rm host,loc}$ along certain line of sight is dependent on the distribution of electrons in the interstellar medium and most electrons within galaxies are produced when the ultraviolet radiation emitted by newly-formed massive stars ionizes surrounding clouds of gas. Such regions of ionized hydrogen are called HII regions. Thus the distribution of electrons within the galaxy is closely related to the distribution of HII regions. HII regions are found in spiral galaxies and irregular galaxies but rarely seen in elliptical galaxies. Observations of HII regions in nearby spiral galaxies reveal  electron densities in the range of $1-10^{2}~{\rm cm}^{-3}$ \cite{Gutierrez2010}. However, the distribution of HII regions is quite inhomogeneous and are concentrated in spiral arms. Other regions in the galaxy typically have a much lower electron density. For instance, within a few hundred parsecs of the Sun, Jenkins \cite{Jenkins2013} find $\langle n_{\rm e}\rangle \sim 0.04 ~{\rm cm}^{-3}$. Thus even in spiral galaxies,  the FRBs may have a low DM$_{\rm host,loc}$ if their origin are not solely confined in HII regions. We are aware that the gas density  as well as star-formation rate density in galaxies increase with increasing redshifts, which suggests that galaxies could have a higher abundance of HII region at high-redshifts though it is difficult to quantify based on current observations. However, the DM$_{\rm IGM}$ increases rapidly with increasing redshifts,
which would largely compensate for the increase in the DM$_{\rm host,loc}$.

Below we examine the possible prospect of FRBs in constraining $w$.
We generate a population of 1000 FRBs in different redshifts between 1 and 3 (the cosmological parameters used in the simulation are $H_0=67.3~{\rm km~s^{-1}~Mpc^{-1}}$, $\Omega_{\rm M}=0.318$, $\Omega_{\rm \Lambda}=0.682$, $\Omega_{\rm b}=0.049$ \cite{Planck2014}). We do not consider the FRBs at lower redshifts since Type Ia SNe \cite{Suzuki2012} and Baryon Oscillation measurements \cite{BOSS2014} have done excellent jobs.
We need $\sim 10^{3}$ FRBs to ``smooth" the fluctuation of the dispersion measures caused by the inhomogeneous IGM (see Fig.\ref{fig:deviation}).
Motivated by a phenomenological fit of the redshift distribution of Gamma-ray Bursts \cite{Shao2011}, we simply assume that the redshift distribution of FRBs follows the Erlang distribution $f(x;k;\lambda)={\lambda^{\rm k} x^{\rm k-1} e^{\rm -\lambda x}}/{\Gamma(k)}$ with the shape parameter $k=2$ and the scale parameter $\lambda=1$.
Since DM$_{\rm sour}$ are assumed to be negligible and ${\rm DM}_{\rm MW}$ is reasonably known, we just simulate DM$_{\rm IGM,obs}+{\rm DM}_{\rm host}$ of each supposed FRB.
At the redshift $z$, the central value of DM$_{\rm IGM,obs}$ is given by eq.(\ref{eq:DM_IGM}) and the random fluctuation is calculated with the $\sigma{\rm [DM]}$ reported in Fig.1 (the model 3) of \cite{McQuinn2014}.
The value and fluctuation of DM$_{\rm\bf host}$ are assumed to take a normal distribution of $N(200, 100^{2})$.

We then divide these 1000 simulated FRBs into 40 redshift bins with a uniform binwidth $\Delta z=0.05$, within each of which we average the DM of the FRBs into $\overline{\rm DM}_{\rm i}$, where $i$ ranging from 1 to 40 is the number of the bin.
In order to get the standard deviation of $\overline{\rm DM}_{\rm i}$ ($\sigma_{\overline{\rm DM}_{\rm i}}$), we repeat the same simulation for the three uncertainties respectively for 10000 times.
Each time we generate 1000 FRBs and put them into the same redshift bins and then obtain the $\overline{\rm{DM}}_{\rm i}$.
Using the same method as Fig.~\ref{fig:deviation} within each bin, we derive the 68\% confidence level of the three components of dispersion of $\overline{\rm DM}_{\rm i}$, i.e. $\sigma_{\overline{\rm DM}_{\rm i,IGM}}$, $\sigma_{\overline{\rm DM}_{\rm i,f_{\rm IGM}}}$ and $\sigma_{\overline{\rm DM}_{\rm i,host}}$.
    We denote $\sigma_{\overline{\rm DM}_{\rm i,IGM}}$ as $\sigma_{\overline{\rm DM}_{\rm i,IGM}}^2$ = $\sigma_{\overline{\rm DM}_{\rm i,IGM}}^2$ + $\sigma_{\overline{\rm DM}_{\rm i,f_{\rm IGM}}}^2$ + $\sigma_{\overline{\rm DM}_{\rm i,host}}^2$.

\begin{figure}
\includegraphics[width=90mm,angle=0]{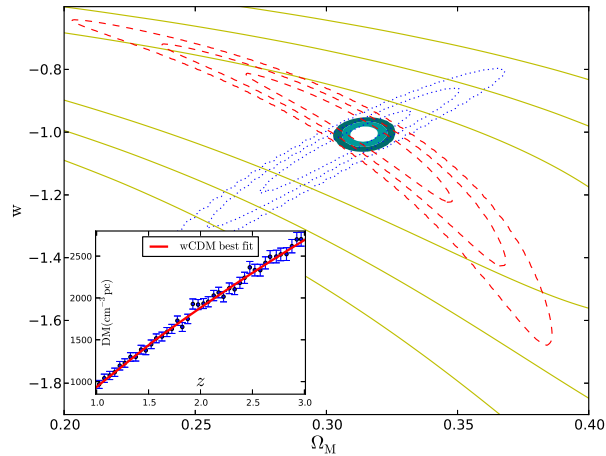}
\caption{The contour lines of constraining $w$ and $\Omega_{\rm M}$ at $1\sigma$, $2\sigma$, and $3\sigma$ confidence level. The solid yellow lines, dotted blue lines and dashed red lines are for 580 SNe Ia, BAO data (consist of real data and forecasted data) and FRBs respectively.
    While the shaded contours are the result combined of the aforementioned three contours.
    We note that FRB and BAO constraints are orthogonal, making this combination of cosmological probes very powerful for investigating the nature of dark energy. The insert is the $\overline{\rm DM}-z$ Diagram of the 1000 simulated FRBs in 40 bins.
}
\label{fig:fit}
\end{figure}

The simulated $\overline{\rm DM}-z$ diagram (similar to the Hubble diagram) of 1000 FRBs is shown as the  insert in Fig.~\ref{fig:fit}.
Now we have acquired 40 ``binned FRBs" to constrain the dark energy equation.
The likelihood for the cosmological parameters can be determined from a $\chi^2$ statistic, where
\begin{equation}
    \chi^2(\Omega_{\rm M},w) = \sum_i{\frac{(\overline{\rm DM}_{\rm i}-\langle{\rm DM}_{\rm IGM,i}\rangle)^2}{\sigma_{\overline{{\rm DM}}_{\rm i,IGM}}^2+\sigma_{\overline{\rm DM}_{\rm i,f_{\rm IGM}}}^2+\sigma_{\overline{\rm DM}_{\rm i,host}}^2}}.
    \label{eq:chi2}
\end{equation}
We constrain the $w$ parameter using the 580 SNe Ia \cite{Suzuki2012}, the simulated FRBs and BAO data respectively and then using these data together.
The BAO data consist of the SDSS data release 10 and 11 \cite{Anderson2013} and the ``forecasted" data at $z=1.0,~1.5,~2.0,~2.5,~3.0$, adopted from \cite{Seo2007}.

By calculating and minimizing the $\chi^2$ for a wide range of the parameters in eq.(\ref{eq:DM_IGM}) and converting each $\chi^2$ into probability density function, we get the contours, as shown in Fig.\ref{fig:fit}, with which it is straightforward to see how effectively the FRBs could be used as a cosmological tool.
We'd like to caution that such tight constraints are obtained under very optimistic assumptions, i.e., both DM$_{\rm host}$ and DM$_{\rm sour}$ are much smaller than $\langle {\rm DM}_{\rm IGM}\rangle$. The validity of such assumptions will be unambiguously tested in the future if a group of FRBs with host galaxies have been detected. If instead DM$_{\rm host}$ and DM$_{\rm sour}$ are found comparable to $\langle {\rm DM}_{\rm IGM}\rangle$, the cosmological studies with FRBs will be hampered unless proper ways to infer DM$_{\rm host}$ and DM$_{\rm sour}$ are available. It may be not unreasonable to expect that as thousands of FRBs with counterparts/redshift-measurements have been collected in the future, our understanding of the contribution of the host galaxies and the FRB sources to the detected dispersion measures could be revolutionized and their influence on constraining the cosmological parameters might be minimized. The other caution is that in eq.(\ref{eq:chi2}) the covariant matrix is assumed to be diagonal. However if the uncertainty mainly comes from the cosmological fluctuation, there should be
off-diagonal correlations which would weaken the power of constraining the cosmological parameters. Specific technique should be developed to remove the covariance in future cosmological studies with real data of FRBs.

In summary, in the optimistic case that satisfying (i) in each narrow-redshift-bin tens events have been measured; (ii) the most distant FRBs were  at $z\geq 3$; (iii) the contribution of host galaxies and the FRB sources to the detected dispersion measures can be ignored, FRBs could serve as a viable cosmic probe and help constrain cosmological parameters for instance the equation of state of dark energy (see Fig.\ref{fig:fit}). If some of these assumptions are invalid, the using of FRBs as a cosmic probe would be challenged. Though in this work we just discuss FRBs, these requests are likely general for using any other kind of cosmological radio transients, if there are, to reliably measure the physical parameters of the universe.
We note that our main conclusions have been confirmed by \cite{gao2014}, one work finished one month later.

\acknowledgments We thank the anonymous referee for insightful comments and are grateful to T. T. Fang and M. McQuinn for helpful communications. This work was supported in part by 973 Programme of China under grants 2013CB837000 and 2014CB845800, National Natural Science of China under grants 11273063, 11303098 and 11361140349, the Foundation for Distinguished Young Scholars of Jiangsu Province, China (No. BK2012047), and the Chinese Academy of Sciences via the 100 Talents program and the Strategic Priority Research Program (Grant No. XDB09000000).

$^\ast$Corresponding author (yzfan@pmo.ac.cn).

\end{document}